\documentclass[a4paper,11pt]{article}
\usepackage{graphics}
\usepackage{jheppub}
\usepackage{epsfig}
\usepackage{float}

\expandafter\ifx\csname package@font\endcsname\relax\else
 \expandafter\expandafter
 \expandafter\usepackage
 \expandafter\expandafter
 \expandafter{\csname package@font\endcsname}%
\fi

\def\be{\begin{equation}}
\def\ee{\end{equation}}

\def\bea{\begin{eqnarray}}
\def\eea{\end{eqnarray}}

\title{ 130 GeV gamma ray line and enhanced Higgs di-photon rate from Triplet-Singlet extended MSSM}

\author{Tanushree Basak}
\author{and Subhendra Mohanty}
\affiliation{Physical Research Laboratory, Ahmedabad 380009, India.}

\emailAdd{tanu@prl.res.in}
\emailAdd{mohanty@prl.res.in}

\abstract{We propose an economic extension of minimal supersymmetric standard model with a SU(2) singlet and $Y=0$ triplet, 
which can explain (i) the 125 GeV Higgs boson without fine tuning, (ii) the 130 GeV $\gamma$-ray line seen at 
Fermi-LAT, (as well as a second photon line at 114 GeV)(iii) an enhanced Higgs di-photon decay rate seen by ATLAS, 
while being consistent with dark matter relic density and recent XENON 100 exclusion limits on spin-independent 
direct detection cross-section. We obtain the required cross-section of $10^{-27}cm^3s^{-1}$ for the 130 GeV 
$\gamma$-ray flux through the resonant annihilation of dark matter via pseudoscalar triplet Higgs of mass $\sim$260 GeV. 
The dark matter is predominantly bino-higgsino which has 
large couplings with photons (through higgsino) and gives correct relic density (through bino).
We get the enhanced Higgs diphoton decay rate, $R_{\gamma \gamma}\simeq 1.224$ dominantly contributed by the
light chargino-loops, which can account for the reported excess seen in the $h\rightarrow \gamma \gamma$ 
channel by ATLAS.}


\begin{document}
\maketitle
\flushbottom
\section{Introduction}

It is well-known that SUSY is the simplest model from protecting the Higgs mass from large radiative corrections without fine tuning. 
In the minimal supersymmetric standard model (MSSM) \cite{MSSM},
the Higgs mass is close to the Z-boson mass at the tree level, which demands a large radiative correction to raise the Higgs mass
to 125-126 GeV seen at the LHC \cite{Atlas,CMS}. This in turn pushes the squark masses in the TeV range and hence the mixing in the top-stop sector becomes significant.
This raises issues about fine tuning - which is somewhat solved by the so-called Next-to-minimal supersymmetric standard model (NMSSM)
by adding a singlet chiral superfield to MSSM \cite{NMSSM}. But to achieve a tree level Higgs mass close to 125 GeV, we need a
large $\lambda S H_u.H_d$ coupling which borders in the nonperturbative regime of $\lambda$ \cite{King}.
Another popular extension is the triplet-extended MSSM models with a $Y=0$,  SU(2)
 triplet superfield \cite{Espinosa:1991wt,Espinosa:1991gr,DiChiara}, 
where  the tree level contribution to the Higgs mass comes from the $\lambda_2 H_d.T_0 H_d$ term. But, 
\cite{DiChiara} shows that the tree-level Higgs mass can be raised atmost to 113 GeV, which would
still require substantial loop corrections from stops. Other possibilities include models with 
two real triplets ($Y=\pm 1$) and one singlet\cite{Agashe:2011ia} - 
studied with a motivation to solve the $\mu$-problem as well as to obtain  a large
correction to the lightest Higgs mass. But, the analysis of the fermionic sector as well as the dark matter of this 
model is cumbersome.
Recently, in \cite{Basak}, it was shown by adding a hypercharge $Y=0$, SU(2)-triplet and a singlet chiral superfield there is an
extra tree-level contribution to the Higgs mass and it can be raised close to 125 GeV at the tree level. Hence, no large  contributions from
stop loops is needed to get the required Higgs mass which
alleviates the fine tuning problem of fixing the stop mass to a high precision at the GUT scale. Therefore a significant 
improvement of the fine tuning is achieved with respect to MSSM, NMSSM and other triplet-SUSY models. 
In addition,  the model contains a dark matter(DM) candidate of mass $\mathcal{O}(100)$ GeV, with a correct relic abundance.

Recently it has been pointed out \cite{Bringmann:2012vr,Weniger:2012tx,Tempel:2012ey,Bringmann:2012ez} that the analysis of the Fermi-LAT gamma-ray data \cite{Fermi-LAT}
reveals the existence of a peak at around 130 GeV coming from the vicinity of the galactic center. Further, it shows that the interpretation of the gamma ray peak
as due to DM annihilation with mass $129.8 \pm 2.4^{+7}_{-13}$ GeV and annihilation cross-section $\langle\sigma v\rangle_{\gamma \gamma} = (1.27
\pm 0.32^{+0.18}_{-0.28})\times 10^{-27} cm^3 sec^{-1}$ fits the signal well.  Numerous studies have been made to accommodate this 
feature  in terms of DM annhilation in both model-independent way \cite{Hooper130} and in specificifically Standard Model (SM) extended by singlets and triplet  
 \cite{Cline:2012nw,Kyae:2012vi,Wang:2012ts}. After the discovery of the Higgs-like boson around mass window 125-126 GeV, 
there is another intriguing
possibility of a signal beyond SM in the $h\rightarrow \gamma \gamma$ channel. The ratio between the Higgs di-photon decay 
rate observed at LHC and the one expected in the SM is  $R_{\gamma\gamma}=1.65^{+ 0.34}_{-0.30}$ for ATLAS ($m_h=126$ GeV) whereas 
CMS have now fallen down to $R_{\gamma\gamma}=0.78^{+0.28}_{-0.26}$ for $m_h=125$ GeV~ \cite{Moriond-2013,Ellis:higgs}. This channel will be an
important discriminator of models as future LHC data pinpoints this number more precisely. The implications of the modified diphoton decay width in a
generic model independent approach have been discussed in ref. \cite{Carena}. Very recently, a vector Higgs-portal 
dark matter model (SM extended by $U(1)_x$ gauge symmetry) \cite{Choi:2013eua} has addressed both Fermi-gamma ray line and diphoton 
excess simultaneously.

In MSSM, the neutralino LSP, being the favourite candidate for DM, annihilates into two photons via loop-suppressed
processes \cite{Bern,Bergstrom} - the cross-section for which is usually too small to explain the signal. But, with a 
bino-like LSP \cite{Kumar} and through the exchange of light slepton and sneutrino the observed 
$\sigma v_{\gamma \gamma}$ is achieved in MSSM. An alternate possibility is
to incorporate the internal Bremsstrahlung (IB), which can also give sharp spectral features in the $\gamma-$ray 
spectrum \cite{IB}. In bino DM annihilation to final state fermions, the fermion mass suppression in the cross section 
is avoided if there is a final state photon with the fermion pair \cite{Bern, Bergstrom}.
In ref.\cite{Shakya} it was pointed out that a significant higgsino component in the DM would  lead to a continuum 
gamma ray spectrum from $W^\pm$ final states and would not be able to explain the gamma ray peak. To avoid this, 
IB from bino dominated LSP's is more promising but there is a problem in getting a natural SUSY model with 
130 GeV bino DM which gives the correct the relic abundance. MSSM could accommodate the enhancement in the di-photon 
decay rate with highly mixed light staus and large $\tan\beta$ \cite{Carena:2012mw}.

In addressing the problem of explaining the 130 GeV gamma ray features, NMSSM models are most widely studied \cite{Das,Kang:2012bq,Chalons}.
In NMSSM, the neutralino DM($\sim 130 $ GeV) annihilates into two photon via resonant channel through psedoscalar singlet
Higgs ($m_{A_s}\sim 260 $GeV) and light charged particle loops. NMSSM can also successfully account for the excess seen in the
$h\rightarrow \gamma \gamma$ channels \cite{NMSSM:diphoton}, in the case of strong singlet-doublet mixing, although the partial width of
 $h \rightarrow b\bar b$ is highly reduced in these models. In a generalised version of NMSSM model(GNMSSM) \cite{SchmidtHoberg:2012ip} 
 simultaneously both the signals from Fermi and LHC has been explained in the same benchmark scenario.

Enhancement of diphoton decay width has been studied well in the triplet extended SUSY models \cite{DiChiara,Delgado,Kang,Wang}, where the
contributions from to the charginos and charged Higgs(triplet like, with large triplet coupling) are taken into account. 
But, so far no benchmark points have been found which at the same time provide a  viable DM in triplet extended SUSY models.

In the present paper, we attempt to explain the 130 GeV gamma ray spectral feature in the triplet-singlet extended MSSM \cite{Basak} through
the resonant annihilation of neutralino LSP into photons via pseudoscalar triplet Higgs of mass $\sim 2m_{DM}$, which couples
to the DM via the Yukawa term, $\lambda_2 T_0 \tilde{H_u^0}.\tilde{H_d^0}$. In addition, our model predicts a second photon peak at around 
114 GeV with the cross-section being 0.75 times $\langle\sigma v\rangle_{\gamma \gamma}$. This DM has a correct relic abundance of 0.109 where dominant
contribution comes from $\langle\sigma v\rangle_{W^+W^-}$. The spin-independent direct detection cross-section is well-below the 
latest XENON100  \cite{xenon100} exclusion limits. Another motivation of this work is to provide an enhanced diphoton decay rate 
compared to SM through the additional contribution from
the light chargino loops. This would be a specific prediction of our model and can be tested in the future collider search.

This article is organised as follows: In section II, the model is described briefly mentioning the details about
the superpotential, bound on the lightest Higgs mass and the fermionic sector. In the next section, we attempt to provide
an explanation for the Fermi-LAT monochromatic gamma ray line features with a neutralino LSP pair annihilation into 
two photon via pseudoscalar Higgs triplet near resonance. We substantiate our claim with a specific 
benchmark scenario which satisfy all desired phenomenological requirements. Section III, shows a detail formulation 
of the diphoton Higgs decay width. We present a short summary and conclusions in the last section.


\section{The Model}

By taking naturalness of the Higgs mass as a guiding criterion, we extend the superpotential of MSSM \cite{Basak}
by adding a SU(2) singlet and triplet chiral superfield S and $T_0$ respectively, where $T_0$ has hypercharge $Y=0$,
\begin{equation}
 \hat{T}_{0}=\begin{pmatrix}
                \frac{\hat{T}^{0}}{\sqrt{2}} & -\hat{T}^{+}_{0}\\
                \hat{T}^{-}_{0} &  \frac{-\hat{T}^{0}}{\sqrt{2}}
               \end{pmatrix} 
\end{equation}
 The most general form of the superpotential can be written as,
\begin{eqnarray}
{\mathcal W}&=&(\mu + \lambda \hat{S}) \hat{H_{d}}.\hat{H_{u}}+ \frac{\lambda_1}{3} \hat{S}^3+\lambda_2 \hat{H_{d}}.\hat{T_{0}}\hat{H_{u}}+
\lambda_{3} \hat{S}^2 Tr(\hat{T}_0)+\lambda_4 \hat{S}Tr(\hat{T}_{0}\hat{T}_{0})+W_{Yuk.}
\end{eqnarray}
where the Yukawa part is same as in the MSSM. To solve the $\mu$-problem, we reduce the general superpotential to the scale-invariant
form as, which then possess an accidental $Z_3$-symmetry,
\begin{eqnarray}
W_{sc.inv.}&=&\lambda \hat{S}\hat{H_{d}}.\hat{H_{u}}+ \frac{\lambda_1}{3} \hat{S}^3+\lambda_2 \hat{H_{d}}.\hat{T_{0}}\hat{H_{u}}+
\lambda_4 \hat{S}Tr(\hat{T}_{0}\hat{T}_{0})+W_{Yuk.}
\label{W_s}
\end{eqnarray}
Therefore, an effective $\mu$-term is generated when the
neutral components of $S$ and $T_0$ acquire vacuum expectation value (vev) $v_s$ and $v_t$ respectively,
\begin{equation}
 \mu_{eff}= \lambda v_s-\frac{\lambda_2}{\sqrt{2}} v_t
\label{mu}
\end{equation}
Here, $v_{u}^{2}+v_{d}^{2} = v^{2}=(174)^{2} GeV^{2}$ (where, $\langle H_{u}^{0}\rangle=v_{u}$ , 
$\langle H_{d}^{0}\rangle=v_{d}$) and  $\tan\beta= \frac{v_{u}}{v_{d}}$.

Due to the addition of triplet, the $\rho$-parameter deviates from unity by a factor of $4\frac{v_t^2}{v^2}$ at the tree level. The 
present bound on $\rho$-parameter, $\rho = 1.0004^{+0.0003}_{-0.0004}$, 
poses strong constraint on the triplet vev $v_t$ from the Electroweak (EW) precision tests such that, $v_t\leq 4$ GeV \cite{Beringer:1900zz} 
at 95$\%$ C.L.

The scalar potential of this model consists of three parts,
\begin{eqnarray}
V&=& V_{SB} + V_{F} + V_{D}
\label{scalarV}
\end{eqnarray}
where, $V_{SB}$ consists of the soft-supersymmetry breaking term associated with the superpotential in equation(\ref{W_s}),
\begin{eqnarray}
 V_{SB}&=& m_{H_{u}}^{2}[\lvert H_{u}^{0}\rvert^{2}+\lvert H_{u}^{+}\rvert^{2}] + m_{H_{d}}^{2}[\lvert H_{d}^{0}\rvert^{2}
+\lvert H_{d}^{-}\rvert^{2}]+m_S^2 \lvert S\rvert^{2}+ m_{T}^{2}Tr(T_0^\dag T_0)+\nonumber \\
&& (-\lambda A_{\lambda} S H_u. H_d +\frac{\lambda_1}{3}A_{\lambda_1} S^3 +\lambda_2 A_{\lambda_2} H_{d}.T_0H_{u}+
\lambda_{4} B_{\lambda}S Tr(T_0^2)+h.c)
\end{eqnarray}
$V_{F}$ and $V_{D}$ are the supersymmetric potential derived from F-terms and D-terms \cite{Basak} respectively.

The CP-even higgs sector consists of four massive higgs as h, $H_1$, $H_2$ and $H_3$. Scalar parts of the singlet
and triplet contribute significantly in the enhancement of the bound \cite{Espinosa:1991gr} on the lightest physical Higgs mass
at the tree level as,
\begin{eqnarray}
m_{h}^{2}\leqslant M_Z^2\left[\cos^2 2\beta+\frac{2\lambda ^2}{g_1^2+g_2^2}\sin^2 2\beta+\frac{\lambda_2^2}{g_1^2+g_2^2}\sin^2 2\beta \right]
\label{bound}
\end{eqnarray}
For moderate values of $\lambda$ and $\lambda_2$, the tree level mass can be lifted so that no large radiative corrections from
the stop sectors are required to obtain $m_h=125-126$ GeV. This indeed reduces the fine-tuning for the EW scale compared to MSSM, NMSSM
and other triplet extended SUSY models. 

The CP-odd higgs sector contains three pseudo-scalar Higgs $A_1$, $A_2$ and $A_3$.
It always contains a Goldstone mode $G^0$, which gives mass to Z-boson, and can be written as,
\begin{equation}
 G^0 = \cos\beta H_{d_I}^0-\sin\beta H_{u_I}^0 \nonumber
\label{goldstone}
\end{equation}
Likewise, there are three massive charged higgs $H_1^\pm$, $H_2^\pm$ and $H_3^\pm$ and the Goldstone mode $G^\pm$
gives mass to the W-bosons.

In the fermionic sector, the neutral component of the triplet and singlet i.e, $\tilde{T^0}$ and $\tilde S$ mix with the
higgsinos and the gauginos.The neutralino mass matrix, in the gauge basis
($\tilde{B},\tilde{W^{0}},\tilde{H_{d}^{0}},\tilde{H_{u}^{0}},\tilde{S},\tilde{T^0}$) reads,
\begin{equation}
\mathcal{M}_{\bar{G}} = \begin{pmatrix}
  M_{1}&  0&  -c_{\beta}s_{w}M_Z&  s_{\beta}s_{w}M_Z&  0&  0\\
  0&  M_{2}&  c_{\beta}c_{w}M_Z&  -s_{\beta}c_{w}M_Z&  0&  0\\
 -c_{\beta}s_{w}M_Z&  c_{\beta}c_{w}M_Z&  0& -\mu_{eff}&  -\lambda v_{u}& \frac{\lambda_{2}}{\sqrt{2}}v_u\\
  s_{\beta}s_{w}M_Z& -s_{\beta}c_{w}M_Z& -\mu_{eff}& 0& -\lambda v_{d}& \frac{\lambda_{2}}{\sqrt{2}}v_d\\
  0&  0& -\lambda v_{u}&  -\lambda v_{d}& 2\lambda_1 v_s& 2\lambda_4 v_t\\
  0&  0& \frac{\lambda_{2}}{\sqrt{2}}v_u&  \frac{\lambda_{2}}{\sqrt{2}}v_d& 2\lambda_4 v_t& 2\lambda_4 v_s
 \end{pmatrix}
\end{equation}
where, $M_1$, $M_2$ are the soft breaking mass for Bino and Wino respectively. The lightest neutralino $\tilde{\chi}_1^0$,
being the lightest supersymmetric particle (LSP), turns out to be a viable dark matter(DM) candidate.

 Similarly, the charged component of the triplet, $\tilde{T}^+$ and $\tilde{T}^-$
contribute to the chargino mass matrix.
The chargino matrix in the gauge basis $\tilde{G}^+$ and $\tilde{G}^-$ is given by,
\begin{equation}
 \mathcal{M}_{ch} = \begin{pmatrix}
        M_2& \frac{1}{\sqrt{2}}g_2 v_d& g_2v_t\\
        \frac{1}{\sqrt{2}}g_2 v_u& \lambda v_s+\frac{\lambda_2}{\sqrt{2}}v_t& \lambda_2 v_d\\
        -g_2v_t& \lambda_2 v_u& 2\lambda_4 v_s
       \end{pmatrix}
\end{equation}
where,
\begin{center}
$ \tilde{G}^+ = \begin{pmatrix}
                \tilde{W}^+\\
                \tilde{H_u}^+\\
                \tilde{T}^+
               \end{pmatrix}$ , $ \tilde{G}^- = \begin{pmatrix}
                \tilde{W}^-\\
                \tilde{H_d}^-\\
                \tilde{T}^-
               \end{pmatrix}$
\end{center}
Since, $\mathcal{M}_{ch}^T\neq \mathcal{M}_{ch}$, this matrix is diagonalised via bi-unitary transformation, which requires two 
distinct unitary matrices U and V such that,
\begin{eqnarray}
 \tilde{\chi}^+ &=& V \tilde{G}^+, \nonumber \\
 \tilde{\chi}^- &=& U \tilde{G}^-
\end{eqnarray}
The diagonal matrix reads,
\begin{eqnarray}
 U^*\mathcal{M}_{ch}V^{-1}&=&\begin{pmatrix}
        m_{\tilde{\chi}_1^\pm} &0 &0\\
        0 &m_{\tilde{\chi}_2^\pm} &0\\
        0 &0 &m_{\tilde{\chi}_3^\pm}
       \end{pmatrix}
\label{char_diag}
\end{eqnarray}
and similarly the hermitian conjugate of eqn.\ref{char_diag} also gives diagonal chargino mass matrix.


\section{130 GeV Fermi gamma ray line}

In this model, the dark matter is the LSP $\tilde{\chi}_1^0$ which can be expressed in the gauge 
basis as,
\begin{eqnarray}
 \tilde{\chi}_1^0&=&N_{11}\tilde{B}+N_{12}\tilde{W_3^0}+N_{13}\tilde{H_d^0}+N_{14}\tilde{H_u^0}+N_{15}\tilde{S}+N_{16}\tilde{T^0}
\end{eqnarray}
where, $N_{11}^2$ is the bino-fraction, $N_{12}^2$ is the wino-fraction, $N_{13}^2 + N_{14}^2$ is the higgsino-fraction, $N_{15}^2$ and $N_{16}^2$
are the singlino and triplino-fraction respectively.

We scan the corresponding regions of the parameter space of the triplet-singlet model \cite{Basak} and tune the 
couplings and masses, such that they satisfy all desired phenomenological
properties. In Table.1, we show a sample set of benchmark points for a particular choice of $\tan\beta=1.8$ specifying all the parameters,
couplings and soft masses at the EW scale.

\begin{table}[ht!]
\label{table1}
\begin{center}
\begin{tabular}{|c|c|c|} \hline
\multicolumn{2}{|c|} {\bf Parameters at EW scale}  \\\hline
$\tan\beta$ &1.8  \\\hline
$\lambda$  & 0.55\\\hline
$\lambda_1$  & 0.20\\\hline
$\lambda_2$  & 0.80\\\hline
$\lambda_4$  & 0.25\\\hline
$\mu_{eff}$[GeV]  &246\\\hline
$A_\lambda$[GeV] &400\\\hline
$A_{\lambda_1}$[GeV] &-50 \\\hline
$A_{\lambda_2}$[GeV] &297.6 \\\hline
$B_\lambda$[GeV] &270 \\\hline
$v_t$[GeV]  &2 \\\hline
$M_1$[GeV]  &154.5\\\hline
$M_2$[GeV]  &375 \\\hline
\hline
\multicolumn{2}{|c|} {\bf Higgs Spectrum [GeV]} \\\hline
$m_h^{Tree}$ &122.93 \\\hline
$m_{H_1}$    &175.29 \\\hline
$m_{H_2}$  &457.27  \\\hline
$m_{H_3}$  &538.86  \\\hline
$m_{A_1}$  &142.12  \\\hline
$m_{A_2}$  &260.54   \\\hline
$m_{A_3}$  &534.56 \\\hline
$m_{H_1}^\pm$  &133.13   \\\hline
$m_{H_2}^\pm$  &365.61   \\\hline
$m_{H_3}^\pm$  &545.59  \\\hline
\end{tabular}
\hspace*{10mm}
\begin{tabular}{|c|c|c|} \hline
\multicolumn{2}{|c|} {\bf Neutralino Masses [GeV]}    \\\hline
$m_{\tilde{\chi}_1^0}$  &130.02   \\\hline
$m_{\tilde{\chi}_2^0}$  &189.0    \\\hline
$m_{\tilde{\chi}_3^0}$  &215.47   \\\hline
$m_{\tilde{\chi}_4^0}$  &269.30   \\\hline
$m_{\tilde{\chi}_5^0}$  &283.49   \\\hline
$m_{\tilde{\chi}_6^0}$  &414.20   \\\hline
\hline
\multicolumn{2}{|c|}{{\bf Chargino Masses [GeV]}} \\\hline
$m_{\tilde{\chi}_1^{\pm}}$  &131.92    \\\hline
$m_{\tilde{\chi}_2^{\pm}}$  &299.38    \\\hline
$m_{\tilde{\chi}_3^{\pm}}$  &422.24    \\\hline
\hline
\multicolumn{2}{|c|} {\bf Observables}       \\\hline
$\Omega h^2$            &   0.109\\\hline
$\sigma(p)_{SI} [10^{-9}\text{pb}$] & 0.681   \\\hline
$\langle \sigma v\rangle({\chi}^0_1 {\chi}^0_1\to \gamma \gamma)\
[10^{-27}\text{cm}^3\ \text{s}^{-1}]$ & 1.249 \\\hline
$\langle \sigma v\rangle({\chi}^0_1 {\chi}^0_1\to Z \gamma)\
[10^{-27}\text{cm}^3\ \text{s}^{-1}]$ & 0.94 \\\hline
$\langle \sigma v\rangle({\chi}^0_1 {\chi}^0_1\to WW)\
[10^{-27}\text{cm}^3\ \text{s}^{-1}]$ & 3.57\\\hline
$\langle \sigma v\rangle({\chi}^0_1 {\chi}^0_1\to Z Z)\
[10^{-27}\text{cm}^3\ \text{s}^{-1}]$ & 0.62  \\\hline
$\langle \sigma v\rangle({\chi}^0_1 {\chi}^0_1\to b \bar{b})\
[10^{-27}\text{cm}^3\ \text{s}^{-1}]$ & 0.045  \\\hline
$\langle \sigma v\rangle({\chi}^0_1 {\chi}^0_1\to \tau \bar{\tau})\
[10^{-27}\text{cm}^3\ \text{s}^{-1}]$ & 0.082 \\\hline
$R_{\gamma \gamma}$  & 1.24 \\\hline
\end{tabular}
\vspace*{10mm}
\end{center}
\caption{A sample set of benchmark points for $\tan\beta =1.8$ and $M_1=154.5$ GeV. The mass spectrum indicates all masses 
at the tree-level}
\end{table}

\begin{itemize}

 \item As shown in \cite{Basak}, the CP-even physical Higgs boson receives significant contribution from the singlet and triplet through the terms
$\lambda \hat{S}\hat{H_{d}}.\hat{H_{u}}$ and $\lambda_2 \hat{H_{d}}.\hat{T_{0}}\hat{H_{u}}$ and thus its mass is raised to 122.9 GeV
at tree level. It requires a little contribution from the radiative corrections raise
it to 126 GeV. This lightest CP-even Higgs is SM-like with large $H_u^0$ and $H_d^0$ component.

\item A dominantly triplet-like pseudoscalar Higgs $A_T$ with mass $\sim 260.54$ GeV can be obtained by adjusting the 
soft-trilinear couplings. The psedoscalar triplet $A_T$
has no tree-level coupling with the SM fermions or Z-boson. It can interact with the neutralinos and charginos via the Yukawa
term in the lagrangian like $\lambda_2 A_T \tilde{H_u^0}.\tilde{H_d^0} $. Although the doublet-triplet mixing terms 
like $\frac{\lambda_2^2}{2}\Big[|H_u^0|^2 + |H_d^0|^2 \Big]|T^0|^2$ is present 
in the scalar potential, but $A_T$ cannot decay into two CP-even Higgs boson, $ m_h$. Therefore the width of $A_T$ is small, i.e,
$\Gamma_T\simeq 6.84$ MeV- which boosts the Breit-Weigner propagator and cross-section $\langle\sigma v\rangle_{\gamma \gamma}$.

\begin{figure}[t!]
\vspace*{10 mm}
\begin{center}
\includegraphics[scale=0.6]{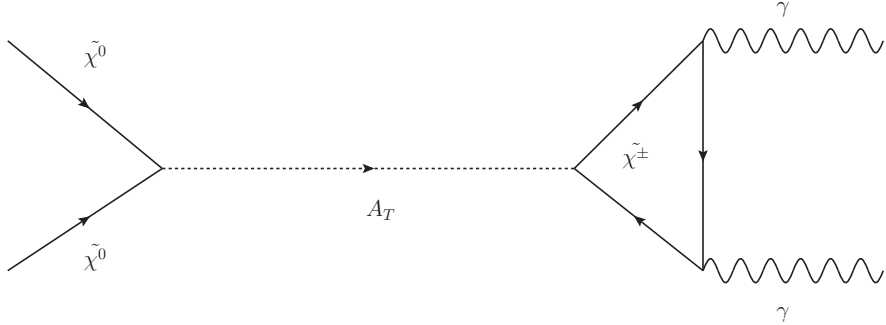}
\vspace*{3mm}
\caption{The dominant diagram for the resonant pair annihilation of neutralino into two photons via psedoscalar triplet Higgs $A_T$}
\label{fig1}
\end{center}
\end{figure}

\item The LSP $\tilde{\chi}_1^0$ is dominantly bino-like ($N_{11}\sim 0.84$) but contains substantial higgsino-fraction 
($N_{13}\sim -0.31$ and $N_{14}\sim 0.36$).
By suitably tuning the soft masses $M_1$ and $M_2$ , the desired mass of 130 GeV is obtained.Varying $M_1$ between 150-160 GeV, we
obtain $127\leq M_{\tilde{\chi}^0}[\text{GeV}]\leq 133$. Here, $\mu$-eff$\sim 246$ GeV being less than $v_s\sim 450$ GeV makes the singlino
 ($N_{15}\sim -0.19$) and triplino-fraction ($N_{16}\sim 0.10$) 
less in $\tilde{\chi}_1^0$. Again, since $M_1$ is lighter than $\mu$-eff, we get a enhancement in the bino fraction compared to
higgsino. But, the significant higgsino fraction is required to get large value of $\langle\sigma v\rangle_{\gamma \gamma}$ through
the resonant annihilation via psedoscalar Higgs $A_T$ and the light chargino loops. In FIG.\ref{fig1} the resonant annihilation channel into
two photon is shown. The lightest chargino $\tilde{\chi}_1^+$ and the DM are
almost degenarate and is also dominantly higgsino-like. 

The pair annihilation of $\tilde{\chi}^0$, with mass $129.8 \pm 2.4^{+7}_{-13}$ GeV into two photon demands a cross-section of
$\langle\sigma v\rangle_{\gamma \gamma} = (1.27\pm 0.32^{+0.18}_{-0.28})\times 10^{-27} cm^3 sec^{-1}$ in order to fit the 
Fermi-LAT signal \cite{Fermi-LAT}.

A simplified form of the analytical expression of $\langle\sigma v\rangle_{\gamma \gamma}$ following \cite{Hooper130} ,
\begin{equation}
 \langle\sigma v\rangle_{\gamma \gamma} = \frac{\alpha^2 g_f^2 g_\chi^2}{256\pi^3}\frac{m_{\chi^+_1}^2}{[(4m_{DM}^2-m_{A_T}^2)^2+
\Gamma_T^2 m_{A_T}^2]}\times [arctan[(m_{\chi^+_1}^2-m_{DM}^2)/m_{DM}^2]^{-1/2}]^2
\end{equation}
where, $g_\chi$ and $g_f$ are the couplings of psedoscalar Higgs $A_T$ with DM and the charged fermion in the loop respectively. Here,
we take the assumption that only the lightest chargino,with mass 131.9 GeV contributes significantly. Upto a crude approximation,
$g_\chi \sim \lambda_2 N_{13} N_{14}$ and $g_f\sim \lambda_2 U_{12}V_{12}$, where U and V are diagonalising matrix for the charginos.
Finally, in the resonance limit of $m_{A_T}\sim 2 m_{DM}$
and $m_{\chi^+}\rightarrow m_{DM}$, the pair annihilation cross-section becomes $\sim 1.249\times 10^{-27}cm^3 s^{-1}$. 
However, the mass of the triplet-like CP-odd scalar Higgs has to lie accidentally close to 260 GeV to a precision $\le 1.5$ GeV. 
FIG.\ref{fig2} shows the behaviour of $\sigma v_{\gamma \gamma}$ with the psedoscalar triplet mass near resonance, 
this clarifies the need of tuning of both $M_{\tilde{\chi}_1^0}$ and $m_{A_T}$.

\begin{figure}[t!]
\vspace*{10 mm}
\begin{center}
\includegraphics[scale=1.0]{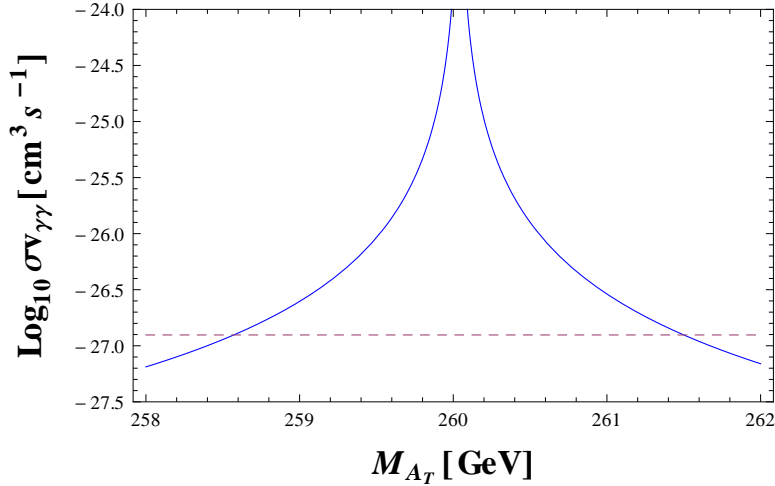}
\vspace*{3mm}
\caption{Plot of $\sigma v_{\gamma \gamma} $ as a function of psedoscalar mass $M_{A_T}$. The dashed line shows the maximum value of
$\langle\sigma v\rangle_{\gamma \gamma}\simeq 1.249\times 10^{-27}cm^3 s^{-1}$. }
\label{fig2}
\end{center}
\end{figure}

\item {\bf A second $\gamma$-ray line at 114 GeV :}
Apart from the monochromatic $\gamma$-ray line at 130 GeV, there is another intriguing hint for a second line at $\sim$111 GeV
 \cite{Rajaraman:2012db,Su}, where the best fit to the relative cross-section is $\langle \sigma v\rangle_{\gamma Z}/
\langle \sigma v\rangle_{\gamma \gamma}=0.66^{+0.71}_{-0.48}$ \cite{Bringmann:2012ez}. A second photon line at 114 GeV is expected 
from kinematics if there is a $Z\gamma$ final state in the annhilation of $\tilde {\chi}_1^0$, 
\begin{equation}
 E_{\gamma}= m_{\tilde {\chi_1^0}}(1-\frac{m_Z^2}{4m_{\tilde {\chi_1^0}}^2})
\end{equation}
where, $E_{\gamma}=114$ GeV for $m_{\tilde {\chi}_1^0}=130$ GeV. The cross-section for $\langle \sigma v\rangle_{\gamma Z}$ is 
calculated using an approximation of the formulae given in \cite{Ullio:1997ke}. Here, we find that for the set of benchmark 
points presented in Table.I, $\langle \sigma v\rangle_{\gamma Z}\simeq 0.943\times 10^{-27}cm^3 s^{-1}$.

\end{itemize}


{\it{\bf Relic Density :}}\\ Another issue with dark matter is to satisfy the correct relic abundance, which is difficult in case 
when it is dominantly higgsino-like since it couples to gauge boson very efficiently and thus leads to large pair annihilation 
cross-section. This kind of interaction can be reduced by an enhanced bino component. We find a neutralino DM with $N_{11}\sim 0.84, 
N_{13}\sim -0.31$ and $N_{14}\sim 0.36$, which makes the relic density 0.109. The pair annihilations into final state $W^+W^-$, ZZ, 
$b\bar{b}$, $\tau^+\tau^-$ are shown in Table.I, calculated using micrOMEGAs2.4 \cite{micromegas}.
Thus, a bino dominated but with a substantial higgsino component dark matter is preferable in order 
to satisfy the latest PLANCK result, i.e, $\Omega_{\chi} h^2=0.1199 \pm 0.0027$ at $68\%$ CL \cite{planck} whereas the 
corresponding value from the 9-year WMAP data is $\Omega_{\chi} h^2 = 0.1148 \pm 0.0019$ \cite{wmap9}. 

\vspace{0.5cm}


{\it{\bf Calculation of spin-independent cross-section : }}\\
Starting from a low-energy neutralino-quark effective lagrangian for spin-independent interaction,
\begin{equation}
 L_{eff}=a_q \bar {\tilde{\chi}}_1^0\tilde{\chi}_1^0 \bar{q}q
\end{equation}
where, $a_q$ is the neutralino-quark coupling, we obtain 
the scattering cross section (spin-independent) for the dark matter off of a proton or neutron as,
\begin{equation}
 \sigma_{scalar}=\frac{4m_r^2}{\pi}f_{p,n}^2
\end{equation}
where, $m_r$ is the reduced mass of the nucleon and $f_{p,n}$ is the neutralino coupling to
proton or neutron\cite{Jungman:1995df,Bertone:2004pz}, given by
\begin{equation}
\label{scalarterms}
f_{p,n} = \sum_{q=u,d,s}  f_{Tq}^{(p,n)} a_q  \frac{m_{p,n}}{m_q}  + \frac{2}{27}f_{TG}^{(p,n)}
\sum_{q=c,b,t} a_q \frac{m_{p,n}}{m_q},
\end{equation}
where $f_{Tu}^{(p)}=0.020 \pm 0.004, f_{Td}^{(p)}=0.026 \pm 0.005, f_{Ts}^{(p)}=0.118 \pm 0.062,
f_{Tu}^{(n)}=0.014 \pm 0.003, f_{Td}^{(n)}=0.036 \pm 0.008$ and $f_{Ts}^{(n)}=0.118 \pm 0.062$
\cite{fvalues}. $f_{TG}^{(p,n)}$ is related to these values by
\begin{equation}
f_{TG}^{(p,n)} = 1 - \sum_{q=u,d,s} f_{Tq}^{(p,n)}.
\end{equation}
In deriving an approximate form of $a_q/m_q$  we ignore contributions from the squark exchange diagrams because of the latest LHC 
bounds on squark masses \cite{ATLAS:2012ona,Chatrchyan:2012lia}. Thus, $a_q$ receives significant contribution  
from the t-channel exchange of CP-even Higgs bosons. The analytical form of $a_q$ goes roughly as,
\begin{equation}
 \frac{a_q}{m_q} \simeq \frac{S_{\chi \chi h_i}}{m_{h_i}^2}  S_{h_i qq} 
\end{equation}
where, $S_{\chi \chi h_i}$ is the coupling between the neutralino and the CP-even Higgs bosons. For, up-type quarks, 
$S_{h_i uu} = \frac{g_2}{2M_w \sin\beta}S_{i1}$ and down-type, $S_{h_i dd} = \frac{g_2}{2M_w \cos\beta}S_{i2}$. Now, the coupling 
$S_{\chi \chi h_i}$ is a product of different combinations of $\lambda$'s, $N_{1k}$ and $S_{i1,2}$. $S_{ij}$ is the matrix 
which diagonalises the CP-even Higgs matrix, and the weak eigenstate basis is $(H_{u_R}^0 , H_{d_R}^0 , T_R^0 , S_R)$. 
$N_{1k}$'s are the different components of the lightest neutralino dark matter. 
Under the assumption that only the lightest physical Higgs boson, i.e, $h_1$ ($m_{h_1}\simeq 125.8$ GeV) 
contributes dominantly, $S_{\chi \chi h_1}$ takes the form, 
\begin{eqnarray}
 S_{\chi \chi h_1} &\simeq & g_2(N_{12}- \tan\theta_W N_{11})(S_{11}N_{13}-S_{12}N_{14})\nonumber \\
&& -\sqrt{2}\lambda(S_{11}N_{14}N_{15}+S_{12}N_{13}N_{15}+S_{14}N_{14}N_{13})+\sqrt{2}\lambda_1 S_{14}N_{15}^2  \nonumber \\
 && +\lambda_2 (S_{11}N_{16}N_{13}+S_{12}N_{16}N_{14}+S_{13}N_{13}N_{14})\nonumber \\
&& + \sqrt{2}\lambda_4(S_{14}N_{16}^2+2S_{13}N_{15}N_{16})
\end{eqnarray}
where the first term is the usual MSSM contribution, the second and third terms are due to the singlet. 
The fourth and fifth terms are the triplet contribution coming from $\lambda_2H_dT_0H_u$ and 
$\lambda_4 STr(T_0T_0)$ in the superpotential respectively. Numerical values of the components $S_{1j} (j=1,..,4)$ as 
obtained from the benchmark point are, $S_{11}\sim 0.885$, $S_{12}\sim 0.463$, $S_{13}\sim0.026$ and $S_{14}\sim-0.037$.
In this model, we find that the spin-independent cross-section $\sigma_p\simeq 6.8\times 10^{-10}$ pb, which is well below the upper bound presented by 
the latest XENON 100 results \cite{xenon100} and can be accessible by the future XENON 1T experiment.


\section{Di-photon Higgs decay rate}

In the SM, the diphoton decay of the Higgs boson is attributed through the W-boson loop
and the contribution from the top-quark destructively interferes with the dominant W-boson contribution.
 The analytic expression for the diphoton partial width given as \cite{Ellis:1975ap, Shifman:1979eb}
\be
\Gamma(h\to \gamma\gamma)=\frac{G_F \alpha^2 m_h^3}{128\sqrt{2}\pi^3}\left|A_1(\tau_W)+ N_c Q_t^2  A_{1/2}(\tau_t) \right |^2 \ ,
\ee
where $G_F$ is the Fermi constant, $N_c=3$ is the number of color, $Q_t=+2/3$ is the top quark electric charge in
units of $|e|$, and $\tau_i\equiv 4m_i^2/m_h^2$, $i=t, W$. The loop functions $A_1(\tau_W)$ and $A_{1/2}(\tau_t)$
for spin-1 ($W$ boson) and spin-1/2 (top quark)  particles are given in \cite{Djouadi:2005gj}. The numerical values of the
loop functions for $m_h=125$ GeV are,
\begin{center}
 $A_1(\tau_W)\simeq -8.3$ , $A_{1/2}(\tau_t)\simeq 1.4$
\end{center}
But in SUSY, we have additional contributions from the s-tops and charginos loops, which would significantly interfere
with the SM contributions. Therefore, in general the branching width of Higgs decay to di-photon is formulated as
\cite{Djouadi:2005gj},
\be
\Gamma(h\to \gamma\gamma)=\frac{ \alpha^2 m_h^3}{1024\pi^3}\left|\frac{g_{hVV}}{m_V^2} Q_V^2 A_1(\tau_V)+
\frac{2g_{hf\bar{f}}}{m_f} N_{c,f} Q_f^2  A_{1/2}(\tau_f) +  N_{c,S} Q_S^2 \frac{g_{hSS}}{m_S^2} A_0(\tau_S) \right |^2 \ ,
\ee
In the above the equation $V$, $f$, and $S$ refer to generic spin-1, spin-1/2, and spin-0 particles, respectively.
$Q_V$, $Q_S$  and $Q_f$ are the electric charges of the vectors, scalars and fermions in units of $|e|$, $N_{c,f}$ and
$N_{c,S}$ are the number of fermion and scalar colors. $A_1(\tau_V)$, $A_{1/2}(\tau_f)$ and $A_0(\tau_S)$ are the loop functions
for the vectors, fermions and scalars respectively.

\begin{figure}[t!]
\vspace*{10 mm}
\begin{center}
\begin{tabular}{cc}
\includegraphics[scale=0.8]{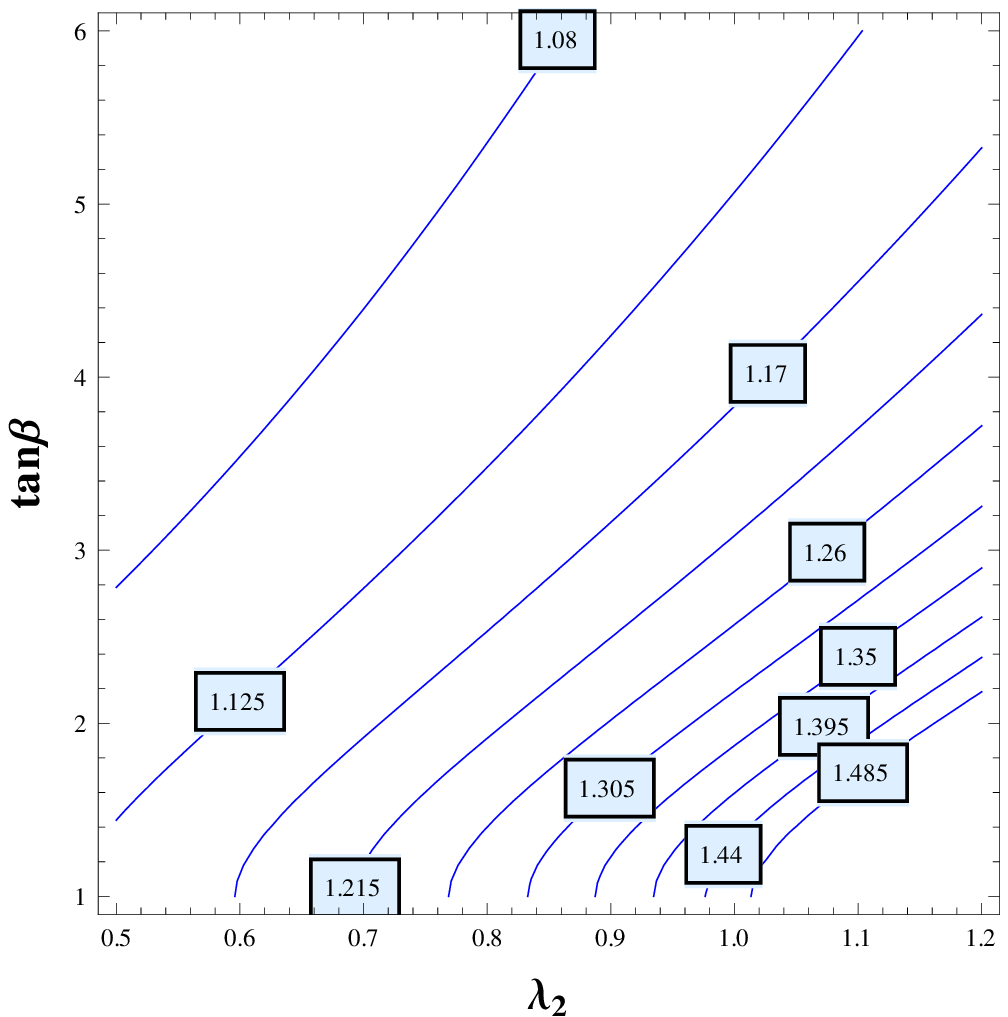}
\ &
\includegraphics[scale=0.8]{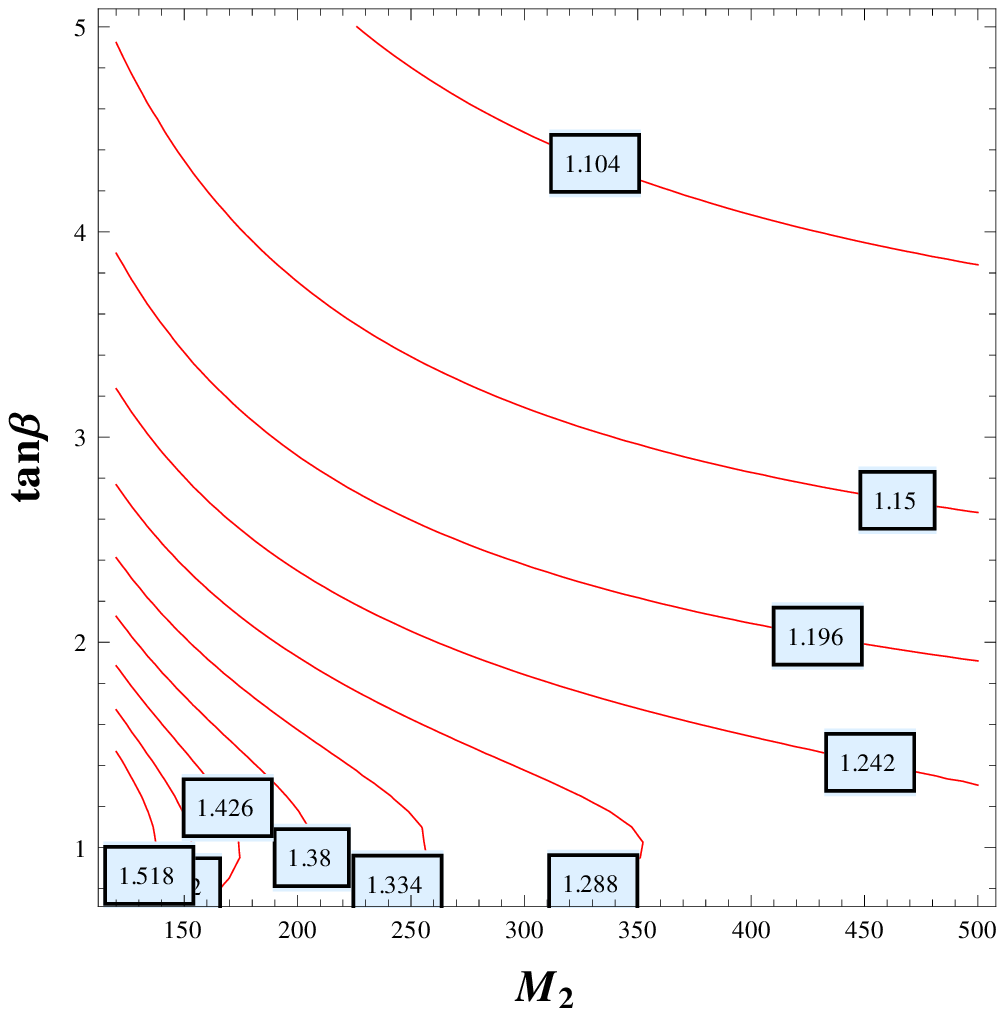}
\end{tabular}
\vspace*{3mm}
\caption{Left Panel : Contours of $R_{\gamma \gamma}$ as a function of $\tan\beta$ and the triplet coupling $\lambda_2$ with 
$M_2=375$ GeV. 
Right Panel : Contours of $R_{\gamma \gamma}$ as a function of $\tan\beta$ and $M_2$ with $\lambda_2=0.8$}
\label{fig3}
\end{center}
\end{figure}

In this model, the additional contribution to the diphoton Higgs decay width comes from the light chargino and the charged Higgs. 
Here, we take the assumption that the lightest charged Higgs (being dominantly triplet-like) only contribute to the decay width, since the other charged 
Higgs are much heavier. Now the term in the potential which gives rise to $hH^\pm H^\pm$ interaction is,
\begin{equation}
 V_F\supset \lambda_2^2 v_u H_u^0 T_0^+ T_0^-
\end{equation}
Therefore, the coupling $g_{hH^\pm H^\pm}$ becomes $\sim \lambda_2^2 v \sin\beta S_{11}C_{13}C_{14}$, where $C_{ij}$ is the 
diagonalising matrix for the charged Higgs and $C_{13}\sim -0.669$ , $C_{14}\sim -0.742$. The loop function for the scalar 
$A_0(\tau_s)$ is given by \cite{Djouadi:2005gj},
\begin{equation}
 A_0(\tau_s)= -\tau_i^2[\tau_s^{-1}-f(\tau_s^{-1})]]
\end{equation}
where, $f(\tau_s)=arc\sin^2\sqrt{\tau_s}$ for, $\tau_s>1$. 

Therefore, considering the main contributions due to charginos, charged triplet, $W$-boson and top quark $t$
and in the limit $m_h^2\ll 4m_{\widetilde \chi_i^{+}}^2$, the diphoton Higgs decay rate with respect to the SM value
becomes \cite{Delgado},
\begin{equation}
R_{\gamma\gamma}=\left|
1+\frac{ {\displaystyle \frac{4}{3}\frac{\partial}{\partial \log v} } \log\det \mathcal M_{ch}(v)
+\frac{g_{hH^\pm H^\pm}}{m_{H_1^\pm}^2}A_0(\tau_s)}{A_1(\tau_W)+{\displaystyle \frac{4}{3} } A_{1/2}(\tau_t)}
\right|^2~,
\label{Rate}
\end{equation}
The numerator (first term) in Eq.~(\ref{Rate}) is given by
\begin{equation}
\frac{\partial}{\partial \log v}  \log\det \mathcal M_{ch}(v)=-\frac{v^2[\sin 2\beta(\lambda_2^2 M_2+2g_2^2\lambda_4v_s)-2\lambda_2g_2^2v_t]}
{2(M_2\lambda_4v_s+g_2^2v_t^2)\mu_{eff}-\frac{1}{2}v^2[\sin 2\beta(\lambda_2^2 M_2+2g_2^2\lambda_4v_s)-2\lambda_2g_2^2v_t]}, 
\end{equation}
and its sign depends on the specific choices for the parameters.
We are specifically interested in the region of parameter space where the numerator is negative (since the denominator is also negative),
such that we obtain, $R_{\gamma \gamma}> 1$. We find that, the factor $g_{hH^\pm H^\pm}/m_{H_1^\pm}^2\sim 0.0024$ 
and thus the contribution due to the extra charged triplet is treated to be negligible compared to the light chargino loops.

We see that for the set of benchmark points specified in Table.I, we obtain chargino masses in the range, $M_{\chi_i^\pm}\ni$
[131.92,299.38,422.24] GeV for $\tan\beta = 1.8$. This choice of parameter gives, $R_{\gamma \gamma}\simeq 1.224$. In FIG.\ref{fig3} 
(left panel), we show
the contours of $R_{\gamma \gamma}$ in the ($\tan\beta,\lambda_2$) plane for $M_2=375$ GeV. We observe that $50\%$ enhancement 
can be  achieved with $\tan\beta\simeq 2$ but the triplet coupling $\lambda_2$ ($\ge 1.1$) then enters into nonperturbative regime. 
Right panel of FIG.\ref{fig3} shows the dependence of $R_{\gamma \gamma}$ on $\tan\beta$ and $M_2$. Here, we note that lowering 
the value of $M_2$ increases the $R_{\gamma \gamma}$, but then we deviate from other phenomenological requirements.

\section{Conclusion}

Recent analysis of the Fermi-LAT data shows existence of a monochromatic $\gamma$-ray line like features at $E_\gamma \sim 130$ GeV in
the vicinity of the galactic center. A possible interpretation comes from DM annihilation into two photons, which demands the annihilation
cross-section to be $1.27\times10^{-27}cm^3 s^{-1}$. We have proposed a triplet-singlet extended MSSM where we obtain the  
lightest CP-even Higgs boson with mass 126 GeV, without much fine-tuning. We scan the parameter space of this model 
and choose a specific set of benchmark points such that it satisfies all phenomenological requirements in order to  
obtain the required cross-section through the pair annihilation of 130 GeV neutralino DM via a psedoscalar Higgs triplet of mass
$M_{A_T}\sim 2 m_{DM}$ near resonance and light chargino loops. The width of the pseudoscalar triplet being small helps in boosting the Breit-Weigner 
cross-section, $\langle\sigma v\rangle_{\gamma\gamma}$. Besides, this model also predicts a second $\gamma$-ray peak at 114
GeV from the annihilation $\chi \chi\rightarrow \gamma Z$, and the cross section is approximately 0.75 times that of
$\langle\sigma v\rangle_{\gamma\gamma}$, which is below the upper limit reported by Fermi LAT.
The dark matter candidate being a mixture of bino-higgsino, leads to a correct 
relic abundance of 0.109, consistent with the PLANCK and WMAP-9 year data. The spin-independent scattering cross-section with nucleons
is $0.68\times10^{-9}$pb, which is well below the latest XENON100 exclusion limits.

Although latest results from CMS seem to favour a SM-like Higgs boson, but on the other hand ATLAS still shows 
a significant excess in diphoton decay width compared to SM as, $R_{\gamma\gamma}=1.65^{+ 0.34}_{-0.30}$ for $m_h=126$ GeV. 
Our model predicts a similar enhancement in the diphoton decay rate as, $R\gamma\gamma \sim 1.224$, 
which is contributed dominantly through the light chargino loops, since the contribution from the extra charged 
triplet is negligible. Such a prediction  opens the possibility of this  model being tested in future LHC runs.


\end{document}